# On The Minimum Mean-Square Estimation Error of the Normalized Sum of Independent Narrowband Waves in the Gaussian Channel

Jacob Binia
New Elective - Engineering Services Ltd
Haifa, Israel
Email: biniaja@netvision.net.il

*Abstract* — The minimum mean-square error of the estimation of a signal where observed from the additive white Gaussian noise (WGN) channel's output, is analyzed. It is assumed that the channel input's signal is composed of a (normalized) sum of N narrowband, mutually independent waves. It is shown that if N goes to infinity, then for any fixed signal energy to noise energy ratio (no matter how big) both the causal minimum mean-square error CMMSE and the non-causal minimum mean-square error MMSE converge to the signal energy at a rate which is proportional to 1/N.

## I. INTRODUCTION

This letter deals with the mean-square error of the optimum estimation of a process $\xi$ where observed from the sum $\eta$

$$\eta(t) = w(t) + \sqrt{q} \int_0^t \xi(s) ds, \quad 0 \le t \le T, \quad (1)$$

where $w$ is a standard Brownian motion independent of $\xi$, and $E_\xi = \int_0^T \xi^2(t) dt < \infty$.

In particular, two examples will be shown for which the causal minimum mean-square error CMMSE and the non-causal minimum mean-square error MMSE exceed $E_\xi$ for each q (no matter how big, but fixed).

In [1] the following example of a non-Gaussian signal was treated

$$\xi_N(t) = \sum_{i=1}^N \sqrt{\frac{2}{T}} \alpha_i \cos(\omega_{k_i} t + \theta_i), \quad (2)$$

where all phases $\theta_1^N$ are uniformly distributed, mutually independent and independent of w, $\omega_{k_i} = \frac{2\pi k_i}{T}$, $i_1^N$, are circular frequencies, and the set $\alpha_1^N$ are positive numbers that fulfill $\sum_{i=1}^N \alpha_i^2 = 1$. In this case for each N we have $E_\xi = 1$.

It was shown in [1] that if signal $\xi$ (1) is replaced by signal $\xi_N$ (2), the following results for the estimation errors $\text{CMMSE}(\xi_N)$ and $\text{MMSE}(\xi_N)$ hold:

$$\text{CMMSE}(\xi_N) = \frac{2}{q} \sum_{i=1}^N \ln(1 + \frac{\alpha_i^2 q}{2}) - \frac{2}{q} D_N(\eta \| \tilde{\eta}), \quad (3)$$

$$\text{MMSE}(\xi_N) = \sum_{i=1}^N \frac{\alpha_i^2}{1 + \frac{\alpha_i^2 q}{2}} - 2 \frac{d}{dq}(D_N(\eta \| \tilde{\eta})). \quad (4)$$

In (3), (4) the divergence $D_N(\eta \| \tilde{\eta})$, where $\tilde{\eta}$ is a Gaussian process with the same covariance function as that of $\eta$, is the following sum

$$D_N(\eta \| \tilde{\eta}) = \sum_{i=1}^N D(\alpha_i^2 q). \quad (5)$$

D(q) is the diminution of the divergence to the case N=1, and is expressed by (see [1])

$$D(q) = \int_0^\infty f(r) \ln(\frac{f(r)}{g(r)}) dr, \quad (6)$$

where f(r) is given by the following Rician probability density function

$$f(r) = r \exp\{-\frac{1}{2}[r^2 + q]\} I_0(r\sqrt{q}), \quad (7)$$

and g(r) is given by the following Rayleigh probability density function

$$g(r) = \frac{r}{(1+q/2)} \exp\{-\frac{1}{2} r^2 /(1+q/2)\}. \quad (8)$$

The solution to integral (6) cannot be expressed in a closed form. Numerical results for the divergence D(q) for some values of q were given in [1].

In section II we will first evaluate the divergence $D(\eta \| \tilde{\eta})$ for *any process* $\xi$ that fulfils (1), for small q. Then, as a corollary, we will express the divergence $D_N(\eta \| \tilde{\eta})$ for the signal $\xi_N$ (that is defined in (2)), assuming that all waves in the sum (2) equally share the same energy, namely $\alpha_i^2 = 1/N, i_1^N$, and N is large. Finally, we will show that the errors $CMMSE(\xi_N)$ and $MMSE(\xi_N)$ tend to 1 (the signal energy) as N goes to infinity, at a rate which is proportional to 1/N.

## II.  MAIN RESULTS

The following lemma holds for *any process* $\xi$ that fulfils (1).

*Lemma1:* As $q \to 0$

$$D(\eta \| \tilde{\eta}) = \frac{1}{2} D^{(2)}(0) q^2 + o(q^2), \qquad (9)$$

where $D^{(2)}(0) = \left. \frac{d^2}{dq^2} D(\eta \| \tilde{\eta}) \right|_{q=0}$ and $o(q^2)$ denotes a function f(q) such that $\lim_{q \to 0} [f(q)/q^2] = 0$.

*Proof*
In order to prove the lemma we recall the following relations between the divergence and the minimum mean-square errors [1]:

$$CMMSE(\tilde{\xi}) - CMMSE(\xi) = \frac{2}{q} D(\eta \| \tilde{\eta}), \qquad (10)$$

$$MMSE(\tilde{\xi}) - MMSE(\xi) = 2 \frac{d}{dq} D(\eta \| \tilde{\eta}). \qquad (11)$$

In (10), (11) $\tilde{\xi}$ is a Gaussian process with the same covariance function as that of $\xi$. Note that the relation between $MMSE(\xi)$, $CMMSE(\xi)$ and the signal-to-noise ratio q ([2], [3]) was used in [1] in order to prove equality (11).

It is clear that as $q \to 0$ the values of $CMMSE(\tilde{\xi})$, $CMMSE(\xi)$, $MMSE(\tilde{\xi})$ and $MMSE(\xi)$ exceed one. Therefore, from (10), (11)

$$D(\xi \| \tilde{\xi})_{q=0} = \frac{d}{dq} (D(\xi \| \tilde{\xi}))_{q=0} = 0. \qquad (12)$$

By Taylor's theorem [4, p. 95] equation (12) implies (9).

*Remark:* In [5] it was shown that if g is a Gaussian variable and z is a random variable which is independent of g and has finite variance, then the following stronger result holds:

$$D(g + q z \| g + q \tilde{z}) = o(q^2).$$

*Corollary:* If $\xi$ in (1) is replaced by $\xi_N$ (that satisfies (2)), then the divergence $D_N(\eta \| \tilde{\eta})$ possesses the following asymptotic behavior as $N \to \infty$

$$D_N(\eta \| \tilde{\eta}) = \frac{1}{2} D^{(2)}(0) \frac{q^2}{N} + o(\frac{q^2}{N^2}). \qquad (13)$$

The corollary follows by adopting (9) to the case of a single narrowband wave (N=1), and by (5) (with $\alpha_i^2 = 1/N, i_1^N$), as follows:

$$D_N(\eta \| \tilde{\eta}) = N D(\frac{q}{N}) = N \frac{1}{2} D^{(2)}(0) (\frac{q}{N})^2 + N o(\frac{q^2}{N^2})$$

$$= \frac{1}{2} D^{(2)}(0) \frac{q^2}{N} + o(\frac{q^2}{N}), \qquad N \to \infty.$$

We can now state our main result.

*LemmaII:* For each fixed q, as $N \to \infty$

$$CMMSE(\xi_N) = 1 - [\frac{1}{4} + D^{(2)}(0)] \frac{q}{N} + o(\frac{q}{N}), \quad (14)$$

$$MMSE(\xi_N) = 1 - [\frac{1}{2} + 2 D^{(2)}(0)] \frac{q}{N} + o(\frac{q}{N}). \quad (15)$$

Lemma II follows from (3), (4) for $\alpha_i^2 = 1/N, i_1^N$ and (13).

Consider now the following signal that is composed from the sum of *Gaussian* narrowband waves

$$g_N(t) = \sum_{i=1}^{N} \sqrt{\frac{1}{T}} (a_{ci} \cos \omega_{k_i} t + a_{si} \sin \omega_{k_i} t), \qquad (16)$$

where $\{a_{ci}, a_{si}\}, i_1^N$ are mutually independent, zero mean Gaussian random variables with variance 1/N, and $\omega_{k_i}, i_1^N$ are as in (2).

Since $\xi_N$ for the case $\alpha_i^2 = 1/N, i_1^N$ and $g_N$ possess same covariance we can consider $g_N$ as a realization of $\tilde{\xi}_N$. In order to evaluate the mean-square errors we can pass through the calculation that was made in [1, section III], or more shortly use the fact that both CMMSE and MMSE seek their maximum for the Gaussian signal. Hence, from (3), (4)

$$CMMSE(g_N) = \frac{2N}{q} \ln(1 + \frac{q}{2N}), \qquad (17)$$

$$MMSE(g_N) = \frac{1}{1 + \frac{q}{2N}}. \qquad (18)$$

Note that even for unlimited (but fixed) q, both CMMSE and MMSE tend to the signal energy (lemma II) as N goes to infinity. While in detection only the energy counts, this is not so in estimation. The inability to estimate the signal of the above example for infinite N follows from two consequences. First, the measure of the normalized sum of the non-Gaussian narrowband waves becomes "more Gaussian" (in the sense of divergence) as N goes to infinity. Hence, the negative parts in (3), (4) vanish. Second, Gaussian processes are impossible to be estimated if all their spectral components are mutually independent and deeply immersed in the noise. In other words, since all spectral components of the signal are independent (such that there is no mutual information between any disjoint sets of components), they have to be estimated separately. However, in this case the individual SNR is low although the total SNR is high.

### III. NUMERICAL RESULTS

In this section we follow the calculation's method from [1] in order to draw the behavior of $D_N(\eta \| \tilde{\eta})$ ((5)-(8)) for q=100, $\alpha_i^2 = 1/N$, $i_1^N$ and $1 \leq N \leq 40$. As we can see from figure 1, the divergence climbs as N increases until it reaches its maximum around N=7 and then decreases.

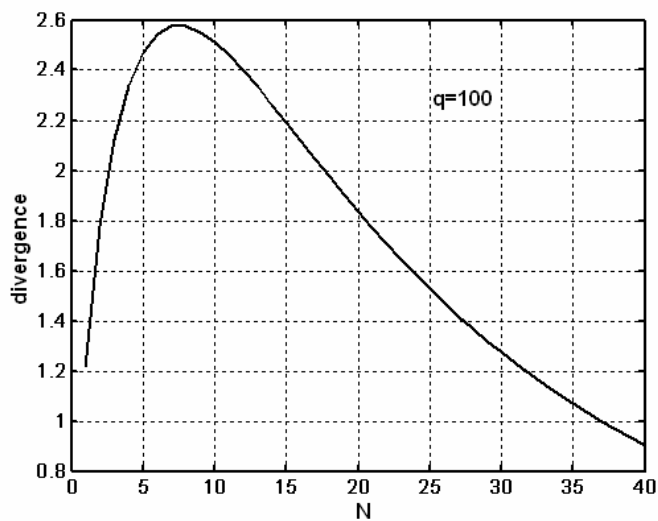

Fig.1: divergence $D_N(\eta \| \tilde{\eta})$ for q=100

Using the divergence results, the minimum error CMMSE was calculated for q=100 and 1≤N≤40. Figure 2 shows plots of $CMMSE(g_N)$ and $CMMSE(\xi_N)$. The difference between them is just the divergence $D_N(\eta \| \tilde{\eta})$ multiplied by 2/q (see (10) and figure 1).

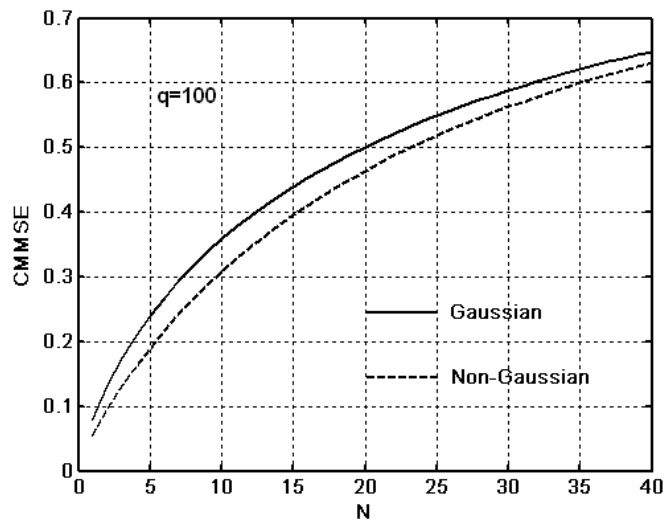

Fig. 2: CMMSE of the normalized sum of N (Gaussian and non-Gaussian) narrowband waves for q=100.